\documentclass[11pt]{article}
\usepackage[T1]{fontenc}
\usepackage[latin1]{inputenc}
\usepackage[dvips]{epsfig}
\usepackage{graphicx}
\usepackage[english]{babel}
\usepackage{amsmath}
\usepackage{amssymb}
\usepackage{amsfonts}
\usepackage{graphicx}
\usepackage{amsfonts}
\title{Geodesic deviation equation in generalized hybrid Metric-Palatini gravity}
\author{~S. Golsanamlou \thanks{%
email:samira.golsanamlou@azaruniv.ac.ir}, K. Atazadeh\thanks{%
email: atazadeh@azaruniv.ac.ir}~ and~M. Mousavi\thanks{%
email: mousavi@azaruniv.ac.ir}\\
{\small Department of Physics, Azarbaijan Shahid Madani University, Tabriz, 53714-161 Iran.}}
\usepackage{color}

\begin{document}
\maketitle

\begin{abstract}
In the context of general relativity, the geodesic deviation equation (GDE) relates the Riemann curvature tensor to the relative acceleration of two neighboring geodesics. In this paper, we consider the GDE for the generalized hybrid Metric-Palatini gravity and apply it in this model to investigate the structure of time-like, space-like, and null geodesics in the homogeneous and isotropic universe. We propose a particular case $f(R,{\cal R})=R+{\cal R}$ to study the numerical behavior of the deviation vector  $\eta(z)$ and the observer area-distance $r_{0}(z)$ with respect to redshift $z$.
Also, we consider the GDE in the framework of the scalar-tensor representation of the generalized hybrid Metric-Palatini gravity {\it i.e.} $f(R, {\cal R} )$, in which the model can be considered as dynamically equivalent to a gravitational theory with two scalar fields.  Finally, we extend our calculations to obtain the modification of the Mattig relation in this model.
\end{abstract}

\section{Introduction}

General relativity is a real scientific theory of gravity developed by Albert Einstein in 1915. Einstein's theory of general relativity (GR) is one of the most successful theories in physics, with a set of simple and beautiful field equations. It is highly consistent with with cosmological observations and has created a new insight into space-time concepts \cite{1}. The mathematical framework of this geometric theory is based on Riemannian geometry, which describes the characteristics of the gravitational field using the space-time curvature tensor. One of the basic equations in this theory is the geodesic deviation equation (GDE), which provides the relationship between the Riemann curvature tensor and the relative acceleration between two nearby test particles. This equation describes the relative motion of free-falling particles to bend toward or away from each other under a gravitational field. The GDE provides a very elegant way to understand the properties of space-time and describe the nature of gravitational forces \cite{2,3}.
In GR, particle motion is described by the curvature of space-time, and the curvature is described by the Riemann curvature tensor. The GDE acts as a force equation, in other words, the concept of force is replaced by geometry, and the path of particles is determined by geodesics instead of by the force equation. In 1933, the GDE was investigated for the first time by Synge, who used the GDE for the geometrical interpretation of Riemann curvature and also to explore the properties of Riemannian spaces, and the properties of space-time with constant curvature \cite{4}.
\\
Even though ordinary GR is a powerful gravitational theory, it is not the final answer to all the cosmological and gravitational issues \cite{1}. Alternative theories have been constructed to generalize the standard cosmology, including modified gravity models \cite{4b,4c}. In the last 10 years, $f(R)$ theories have been studied using the Palatini approach, where the metric and the connection are treated as independent fields, see for example \cite{4d}. The metric formalism in $f(R)$ gravity as described in \cite{4e} in which we vary the action with respect to the metric $g_{\mu\nu}$, can be promoted to the Palatini approach in which we vary the action concerning the metric and the connection \cite{4f}. This continues to form a novel modification of general relativity wherein an $f(\cal{R})$ term is added to the metric Einstein-Hilbert Lagrangian \cite{4h}, and the authors can also go further with a modification like $f\left(R,\cal{R}\right)$, where the gravitational action depends on a general function of both the metric and Palatini curvature scalars that is called generalized hybrid Metric-Palatini gravity \cite{4i}. Of note, it has been reported that it was presented that using the dynamically equivalent scalar-tensor representation causes the theory to pass solar system observational constraints \cite{4j}. Cosmological studies of this hybrid Metric-Palatine gravitational theory were also conducted in \cite{4k}. The authors of \cite{4l} explored the Einstein static universe in this theory, as well.
In the present work, motivated by the fact that the GDE has always been studied in several gravitational theories \cite{4m}, we aim to explore the GDE in the context of generalized hybrid Metric-Palatini theory. In addition, the generalized GDE has been studied in various papers for example, in the context of modified gravity theories, it has been considered in an arbitrary curvature-matter coupling theories, {\it i.e.}, $f(R, L_m)$ gravity \cite{harko}, and has also been studied in $f(R,T)$ gravity \cite{houn}, $f(Q,T)$ gravity \cite{shahab}, Brans-Dicke theory \cite{meraj}, $f(Q)$ gravity \cite{Q}, and the chameleon scalar field model \cite{farhodi}. In \cite{jalal} the authors considered the generalized GDE in the brane world. In \cite{meraj2}, the GDE has been considered in Saez--Ballester theory.
\\
The main target of this work is to systematically use the GDE to consider the geometry of the standard Friedmann-Lema\^{\i}tre- Robertson- Walker (FLRW) universe in the context of generalized hybrid Metric-Palatini theory. In this regard, by considering the GDE for time-like, null and space-like geodesic congruences in FLRW geometries and also obtaining the Raychaudhuri equation, we aim to determine the cosmological time evolution of these models. Also, we consider the generalized GDE for fundamental observers besides the modified \textit{Pirani} equation. We study GDE for null vector fields to extract the null GDE equation and investigate the focusing condition for this model, in which the geodesics experience convergence besides the modified \emph{Mattig} relation.
Moreover, we propose a particular case $f(R,{\cal R})=R+{\cal R}$ in order to study the numerical behavior of the deviation vector $\eta(z)$ and observer area-distance $r_{0}(z)$ as a function of redshift. The existence of a maximum point for $\eta(z)$ and $r_{0}(z)$ at a certain redshift indicates that there were maximum values for deviation vector and the observer area-distance in the past when our universe was experiencing the inflationary regime. After that, the universe exited the inflationary regime and the deviation vector gradually decreased with the increase in $z$.
Therefore, our study characterizes the main geometrical and physical properties of the FLRW space-time using the generalized GDE, thereby demonstrating the utility of this equation in obtaining all the basic geometrical and dynamical results of modified standard cosmology in a unified way.

The paper is organized as follows: In section $2$, we review field equations in hybrid metric-Palatine gravity and its cosmological equations and also we study the GDE for fundamental observers and null vector fields in $f(R,{\cal R})=R+f({\cal R})$ gravity. In section 3, we study the GDE in the scalar-tensor representation of $ f(R,{\cal R}) $ gravity. Finally, we close the paper with conclusions in section 4.

\section{Field equations in hybrid Metric-Palatini gravity}

Generalized gravity models, attempt to provide a suitable alternative for dark energy by using the generalization of gravitational equations. That is, instead of Einstein's equations of general relativity, alternative equations are obtained. So, by solving these equations, and without the need for cosmologists to introduce dark energy, accelerated dynamics for the universe can be obtained. In the cosmological context, $ f(R) $ gravity, as an alternative to dark energy, has been introduced to explain the recent acceleration of the universe. As mentioned in the introduction, the modified GR theory has two approaches to obtaining field equations: the metric approach and the Palatini approach. In metric formalisms, the field equations are obtained by the variation of the action with respect to the metric, and in this case, the affine connections are the functions of the metric. In the Palatini approach, the metric and affine connections are considered as two independent variables. In the metric and Palatini formulation, symmetrical connections are assumed. In the Metric-Palatini formulation, in addition to the independence of metric and connection, the condition of symmetry in connections is absent. In this section, to obtain the field equations, we take the following action \cite{4k}.

\begin{equation}\label{Fbi1}
	S=\frac{1}{2\kappa^{2}}\int d^4{x}\sqrt{-g}[R+{f({\cal R})}]+S_{m}.
\end{equation}
In action (\ref{Fbi1}), $ R $ is the Ricci curvature scalar formed, $ \Gamma^{\lambda}_{\mu\nu} $ is Levi-Civita connection, and $ {\cal R}$ is the Palatini curvature of an independent torsionless connection $ \hat{\Gamma}^{\lambda}_{\mu\nu} $, in analogy with the Palatini approach and also $\kappa^{2}=8\pi
G$. Here $G$, is the Newtonian gravitational constant. Variation of the action (\ref{Fbi1}) with respect to the metric yields

\begin{equation}\label{Fbi2}
	G_{\mu\nu}+F({\cal R} ){\cal R} _{\mu\nu}-\frac{1}{2}f({\cal R})g_{\mu\nu}=\kappa^{2}T_{\mu\nu},
\end{equation}
with the usual definition of the matter stress-energy tensor
\begin{equation}\label{Fbi3}
T_{\mu\nu}=-\frac{2}{\sqrt{-g}}\frac{\delta (\sqrt{-g}~L_{m})}{\delta (g_{\mu\nu})},
\end{equation}
where $L_{m}=L_{m}\left(g^{\mu\nu},\Psi\right)$ is the matter Lagrangian including the minimally coupled matter fields $\psi$ to the metric $g_{\mu\nu}$.
Tracing the field equation gives us
\begin{equation}\label{Fbi4}
2f({\cal R} )-F({\cal R} ){\cal R}=-\kappa^{2}T-R.
\end{equation}
Note that we have
\begin{equation}\label{Fbi5}
F({\cal R}) =\frac{df({\cal R})}{d{\cal R}}.
\end{equation}
By rewriting  equation  (\ref{Fbi2}), the Ricci tensor can be expressed as
\begin{equation}\label{Fbi6}
	R_{\mu\nu}=\kappa^{2}T_{\mu\nu}+\frac{1}{2}g_{\mu\nu}(F({\cal R})+R)-F({\cal R}){\cal R}_{\mu\nu}.
\end{equation}
Using equations (\ref{Fbi2}), (\ref{Fbi4}) and (\ref{Fbi6}), we obtain the hybrid Ricci tensor as
\begin{equation}\label{Fbi7}
  {\cal R}_{\mu\nu}=\frac{1}{1+F({\cal R})}\bigg[\frac{1}{2}g_{\mu\nu}\bigg(R+f({\cal R})-\frac{\Box{F({\cal R})}}{F({\cal R} )}\bigg)+\kappa^{2}T_{\mu\nu}
+\frac{3}{2}\frac{1}{F^{2}({\cal R})}F({\cal R})_{,\mu}F({\cal R})_{,\nu}-\frac{\nabla_{\mu}F({\cal R})_{,\nu}}{F({\cal R})}\bigg].
\end{equation}
and from (\ref{Fbi4}) we get
\begin{equation}\label{Fbi8}
R =-\kappa^{2}T+F({\cal R}){\cal R}-2f({\cal R}).
\end{equation}
Thus far, we have extracted some relations by which we will find the basic feature of the geodesic deviation equation in  $f(R,{\cal R})=R+f({\cal R})$ gravity. In the following section, similar to the previous works \cite{4m} we try to assemble the general form of the right-hand side of the GDE in the context of the generalized hybrid Metric-Palatini gravity.

\subsection{Geodesic deviation equation in $f(R,{\cal R})=R+f({\cal R})$ gravity}

The GDE is one of the basic equations in the theory of general relativity and provides the relationship between the Riemann curvature tensor and the relative acceleration between two test particles.
\begin{equation}\label{Fbi9}
	\frac{D^{2}\eta^{\alpha}}{D\nu^{2}}=-R^{\alpha}_{\,\beta\gamma\delta}V^{\beta}\eta^{\gamma}V^{\delta}.
\end{equation}
This equation describes the relative motion of free-falling particles to bend toward or away from each other under a gravitational field. If we describe the geodesic as $ x^{\alpha}=(\nu,s) $, $R _{\alpha\beta\gamma\delta} $ is the Riemann curvature tensor, and $V^{\alpha}=\frac{dx^{\alpha}}{d\nu}$ is the normalized tangent vector that belongs to the geodesics. In the above equation, $ \eta^{\alpha}=\frac{dx^{\alpha}}{ds} $ denotes the deviation vector of these two adjacent geodesics.\\
Note that in several classes of modified gravity theories, some new terms appear on the right-hand side of the equation (\ref{Fbi9}), mainly due to the presence of couplings between different fields and geometric quantities. It may lead to non-conservation of the energy-momentum tensor of matter and thus to the appearance of an extra-force, see \cite{harko}. However, in the hybrid Metric-Palatini gravity, basically, there is no coupling between the matter fields and geometric quantities, and conservation of the energy-momentum tensor is preserved, thus the standard form of GDE is satisfied. Furthermore, as we know that the universe is isotropic and homogeneous, only the time derivatives of the scalar fields appear, and also in the comoving frame, one has $\eta^{0} = 0$, thus, in the scalar tensor framework of this model that it is studied in the following, we again have the standard form of the GDE.
\\
In general, the Riemann tensor can be decomposed as follows \cite{2,3}
\begin{equation}\label{Fbi10}
R _{\alpha\beta\gamma\delta}=\frac{1}{2}\bigg(g_{\alpha\gamma}R _{\delta\beta}-g_{\alpha\delta}R _{\gamma\beta}+g_{\beta\delta}R _{\gamma\alpha}-g_{\beta\gamma}R _{\delta\alpha}\bigg)-\frac{R }{6}\bigg(g_{\alpha\gamma}g_{\delta\beta}-g_{\alpha\delta}g_{\gamma\beta}\bigg)+C_{\alpha\beta\gamma\delta},
\end{equation}
where $C_{\alpha\beta\gamma\delta}$ is the Weyl tensor.\\
In continuation of our study, we take the standard cosmology model line element, the FLRW universe, as
\begin{equation}\label{Fbi11}
ds^{2}=-dt^{2}+a(t)^{2}\left[\frac{dr^{2}}{1-Kr^{2}}+r^{2}d\theta^{2}+r^{2}sin\theta^{2}d\phi^{2}\right],
\end{equation}
where $a(t)$ is the scale factor, and $K$ denotes the three-dimensional spatial curvature with values $-1$, $0$ and $1$.
The energy momentum tensor can be written in the form of a perfect fluid as
\begin{equation}\label{Fbi12}
	T_{\alpha\beta}=(\rho+P)u_{\alpha}u_{\beta}+P~g_{\alpha\beta},
\end{equation}
where $ \rho $ and $ P $ are the energy density and pressure, respectively. The trace of $T_{\alpha\beta}$ is given by
\begin{equation}\label{Fbi13}
	T=-\rho+3P.
\end{equation}
We know that by redefining  the cosmic time, $t$, to the conformal time $\tau$ by $dt = a(t)d\tau$, the FLRW metric (\ref{Fbi11}) can be rewritten as a form of a conformally flat metric, and according to the conformal invariance property of the Weyl tensor, in the homogeneous and isotropic space-time we can set $C_{\alpha\beta\gamma\delta}=0$. Thus, by using equations  (\ref{Fbi6}), (\ref{Fbi8}), and (\ref{Fbi10}), the hybrid Riemann tensor will be in the following form
\begin{flalign}\label{eq14}
	R _{\alpha\beta\gamma\delta}=&\frac{1}{2}\bigg[ (\frac {2R}{3}+f({\cal R}))(g_{\alpha\gamma}g_{\delta\beta}-g_{\alpha\delta}g_{\gamma\beta})+
\kappa^{2}(T_{\delta\beta}g_{\alpha\gamma}-T_{\gamma\beta}g_{\alpha\delta}+T_{\alpha\gamma}g_{\beta\delta}-T_{\alpha\delta}g_{\beta\gamma})\nonumber\\
& +F({\cal R})[-g_{\alpha\gamma} {\cal R}_{\delta\beta}+g_{\alpha\delta}{\cal R}_{\gamma\beta}-g_{\beta\delta}{\cal R}_{\gamma\alpha}
+g_{\beta\gamma}{\cal R}_{\delta\alpha}] \bigg],
\end{flalign}
By contracting $R ^{\alpha}_{\beta\gamma\delta}$ with $ V^{\beta}\eta^{\gamma}V^{\delta} $, equation (\ref{eq14})  can be written as follows
\begin{flalign}\label{eq15b}
	R ^{\alpha}_{\,\beta\gamma\delta}V^{\beta}\eta^{\gamma}V^{\delta}=&\frac{1}{2(1+F({\cal R} ))}\bigg[  \kappa^{2}(-\delta^{\alpha}_{\delta}T_{\gamma\beta}+\delta^{\alpha}_{\gamma}T_{\delta\beta}+T^{\alpha}_{\gamma}g_{\delta\beta}-T^{\alpha}_{\delta}g_{\gamma\beta})+[(\frac{2-F({\cal R} )}{3}R)+ f({\cal R} )+\square F({\cal R} )]\nonumber\\
		&(\delta^{\alpha}_{\gamma}g_{\delta\beta}-\delta^{\alpha}_{\delta}g_{\beta\gamma})-\frac{3}{2F({\cal R} )}(\delta^{\alpha}_{\gamma}F({\cal R} )_{,\delta}F({\cal R} )_{,\beta}-\delta^{\alpha}_{\delta}F({\cal R} )_{,\gamma}F({\cal R} )_{,\beta}+g_{\delta\beta}F({\cal R} )^{,\alpha}F({\cal R} )_{,\gamma}\nonumber\\
	&-g_{\beta\gamma}F({\cal R} )^{,\alpha}F({\cal R} )_{,\delta})+\delta^{\alpha}_{\gamma}\nabla_{\delta}F({\cal R} )_{,\beta}-\delta^{\alpha}_{\delta}\nabla_{\gamma}F({\cal R} )_{,\beta}+g_{\delta\beta}\nabla_{\gamma}F({\cal R} )^{,\alpha}-g_{\beta\gamma}\nabla_{\delta}F({\cal R} )^{,\alpha}\bigg]V^{\beta}\eta^{\gamma}V^{\delta}.
\end{flalign}
The four-velocity is $u^{\alpha}=\left(1,0,0,0\right)$, therefore, from the orthogonality conditions, we have $E=-V_{\alpha}u^{\alpha}=-V_{0}$, $\eta_{\alpha}u^{\alpha}=\eta_{0}u^{0}=0 $, which means that the deviation vector just has non-vanishing spatial components $\eta^{0}=0$.  Moreover, we have $\eta_{\alpha}V^{\alpha}=\eta_{i}V^{i}$.
Note that the Ricci scalar $R$ in the FLRW space-time is only a function of time, thus by taking equation (\ref{Fbi11}), (\ref{Fbi12}), and (\ref{Fbi13}), we can write the following terms as

\begin{equation}\label{Fbi16}
\delta^{\alpha}_{\gamma}T_{\delta \beta} V^{\beta}\eta^{\gamma}V^{\delta}=\left(\epsilon P+\left(\rho+P\right)E^{2}\right)\eta^{\alpha},
\end{equation}

\begin{equation}\label{Fbi17}
\delta^{\alpha}_{\gamma} F({\cal R})_{,\delta}  F({\cal R})_{,\beta} V^{\beta} \eta^{\gamma} V^{\delta}=f^{''2}\dot{{\cal R}}^{2}E^{2}\eta^{\alpha},
\end{equation}

\begin{equation}\label{Fbi18}
g_{\beta\delta}\nabla_{\gamma}F({\cal R})^{,\alpha} V^{\beta} \eta^{\gamma} V^{\delta}=-\epsilon H F^{'} \dot{{\cal R}} \eta^{\alpha},
\end{equation}

\begin{equation}\label{Fbi19}
\delta^{\alpha}_{\gamma} \nabla_{\delta} F({\cal R})^{,\beta} V^{\beta} \eta^{\gamma} V^{\delta}=\left(E^{2}\left(F^{'}\ddot{{\cal R}}+
F^{''}\dot{{\cal R}}^{2}-H F^{'}\dot{{\cal R}}\right)-\epsilon H F^{'} \dot{{\cal R}}\right)\eta^{\alpha},
\end{equation}

\begin{equation}\label{Fbi20}
\square F({\cal R})=-F^{'} \ddot{{\cal R}} - F^{''} \dot{{\cal R}}^{2}-3H F^{'} \ddot{{\cal R}},
\end{equation}
where $\square = \nabla _{\sigma} \nabla^{\sigma}$, and $F({\cal R})=df({\cal R})/d{\cal R}$. Consequently, the right-hand side of the GDE reduces to

\begin{flalign}\label{eq21}
	R^{\alpha}_{\,\beta\gamma\delta}V^{\beta}\eta^{\gamma}V^{\delta}&=\frac{\eta^{\alpha}}{2(1+F({\cal R}))}\bigg[\epsilon\bigg(\frac{2-F({\cal R})}{3}R+2\kappa^{2}P+ f({\cal R})- F^{\prime\prime}({\cal R})\dot{{\cal R}}^{2}- F^{\prime}({\cal R})\ddot{{\cal R}}-5H F^{\prime}({\cal R})\dot{{\cal R} }\bigg) \nonumber\\
	&+E^{2}\bigg(\kappa^{2}(\rho +P)-\frac{3}{2F({\cal R})}F^{\prime}({\cal R})^{2}\dot{{\cal R}}^{2}+F^{\prime}({\cal R})\ddot{{\cal R}}+F^{\prime\prime}({\cal R})\dot{{\cal R}}^{2}-HF^{\prime}({\cal R})\dot{{\cal R}}\bigg)\bigg],
\end{flalign}
 We define the following terms

\begin{equation}\label{Fbi21b}
\rho_{\textrm{eff}}=\kappa^{2}\rho+\frac{F({\cal R})-2}{6}R-\frac{f({\cal R})}{2}-\frac{3F^{'}({\cal R})^{2}\dot{{\cal R}}^{2}}{2F({\cal R})}+
\frac{3F^{''}({\cal R})\dot{{\cal R}}^{2}}{2}+\frac{3}{2}HF^{'}({\cal R})\dot{{\cal R}}+\frac{3}{2}F^{'}({\cal R})\ddot{{\cal R}},
\end{equation}
and
\begin{equation}\label{Fbi21c}
P_{\textrm{eff}}=\kappa^{2}P+\frac{-F({\cal R})+2}{6}R+\frac{f({\cal R})}{2}-
\frac{F^{''}({\cal R})\dot{{\cal R}}^{2}}{2}-\frac{5}{2}HF^{'}({\cal R})\dot{{\cal R}}-\frac{1}{2}F^{'}({\cal R})\ddot{{\cal R}}.
\end{equation}
As a result, equation  (\ref{eq21}) is reformed into a more compact structure

\begin{flalign}\label{eq21d}
	R^{\alpha}_{\,\beta\gamma\delta}V^{\beta}\eta^{\gamma}V^{\delta}=\frac{\eta^{\alpha}}{2(1+F({\cal R}))}\left[\epsilon\left(2P_{\textrm{eff}}\right)+E^{2}\left(\rho_{\textrm{eff}}+P_{\textrm{eff}}\right)\right],
\end{flalign}
which is the modified \textit{ Pirani} equation. Eventually, the generalized GDE in $f(R, {\cal R})$ gravity is

\begin{flalign}\label{eq22}
\frac{D^{2}\eta^{\alpha}}{D\nu^{2}}=-\frac{\eta^{\alpha}}{2(1+F({\cal R}))}\left[\epsilon\left(2P_{\textrm{eff}}\right)+E^{2}\left(\rho_{\textrm{eff}}+P_{\textrm{eff}}\right)\right].
\end{flalign}
In addition, in a particular GR case, {\it i.e.}, with $f\left(R, {\cal R}\right)=R$, we can obtain the original \textit{pirani} equation and the GDE in GR as

\begin{equation}\label{Fbi23}
\frac{D^{2}\eta^{\alpha}}{D\nu^{2}}=-R^{\alpha}_{\,\beta\gamma\delta}V^{\beta}\eta^{\gamma}V^{\delta}=-\left[\frac{\kappa^{2}\rho}{3}\epsilon +\frac{\kappa^{2}}{2}\left(\rho+P\right)E^{2}\right]\eta^{\alpha}.
\end{equation}
This reduction confirms for the correctness of our calculations. In the next section, we study the details of the GDE for fundamental observers.

\subsection{GDE for fundamental observers}

Here, we consider $ V^{\alpha} $ to be  $ u^{\alpha} $ as the four-velocity, and the affine parameter $ \nu $  is interpreted as the time coordinate $ \nu=t $ which satisfies

\begin{equation}\label{Fbi24}
u^{\alpha} u_{\alpha}=\epsilon~~~~~~~~~~\frac{Du^{\alpha}}{Dt}=u^{\beta}\nabla_{\beta}u^{\alpha}=0.
\end{equation}
In this case we get $\epsilon=-1$. In addition we set the vector field normalization with $E=1$. As a result, the generalized \textit{Pirani} equation reduces to

\begin{flalign}\label{eq25}
R^{\alpha}_{\,\beta\gamma\delta}u^{\beta}\eta^{\gamma}u^{\delta}=&\frac{\eta^{\alpha}}{2(1+F({\cal R}))}\bigg[\kappa^{2}(\rho - P)+\frac{F({\cal R})-2}{3}R-f({\cal R})+2F^{\prime\prime}({\cal R})\dot{{\cal R}}^{2}\nonumber\\
		&+2F^{\prime}({\cal R})\ddot{{\cal R}}+4HF^{\prime}({\cal R})\dot{{\cal R}}-\frac{3}{2F({\cal R})}F^{'}({\cal R})^{2}\dot{{\cal R}}^{2}\bigg]=\frac{\eta^{\alpha}}{2(1+F({\cal R}))}\left[\rho_{\textrm{eff}}-P_{\textrm{eff}}\right].
		\end{flalign}
By putting $ \eta_{\alpha}=le_{\alpha} $ (where $e_{\alpha}$ is propagated parallelly along the cosmic time), we find

\begin{equation}\label{Fbi26}
	\frac{De^{\alpha}}{Dt}=0,
\end{equation}
and
\begin{equation}\label{Fbi27}
	\frac{D^{2}\eta^{\alpha}}{Dt^{2}}=\frac{D^{2}l}{Dt^{2}}e^{\alpha}.	
\end{equation}
Using equations (\ref{Fbi27}) , (\ref{Fbi9}) and  (\ref{eq25}) we will have

\begin{flalign}\label{eq28}
	\frac{D^{2}l^{\alpha}}{Dt^{2}}=&\frac{-1}{1+F({\cal R})}\bigg[\frac{F({\cal R} )R}{6}-\frac{R}{3}-	\frac{f({\cal R})}{2}+F^{\prime\prime}({\cal R})\dot{{\cal R}}^{2} +F^{\prime}({\cal R})\ddot{{\cal R} }+2HF^{\prime}({\cal R})\dot{{\cal R}}\nonumber\\
&+\frac{\kappa^{2}(\rho - P)}{2}-\frac{3}{4F({\cal R})}F^{\prime}({\cal R})^{2}\dot{{\cal R}}^{2}\bigg] l=\frac{-1}{1+F({\cal R})}\left[\rho_{\textrm{eff}}-P_{\textrm{eff}}\right].
 \end{flalign}
If we put $ l=a(t) $, we have

\begin{flalign}\label{eq29}
 \frac{\ddot{a}}{a}=&\frac{1}{1+F({\cal R})}\bigg[\frac{-F({\cal R} )R}{6}+\frac{R}{3}+\frac{f({\cal R})}{2}-F^{\prime\prime}({\cal R})\dot{{\cal R} }^{2} -F^{\prime}({\cal R})\ddot{{\cal R}}-2HF^{\prime}({\cal R})\dot{{\cal R}}\nonumber\\
&+\frac{\kappa^{2}\left(-\rho + P\right)}{2} +\frac{3}{4F({\cal R})}F^{\prime}({\cal R})^{2}\dot{{\cal R}}^{2}\bigg]=\frac{1}{1+F({\cal R})}\left[P_{\textrm{eff}}-\rho_{\textrm{eff}}\right].
 \end{flalign}

In order to check the correctness of our result, we should compare the above relation with the result found by building the standard modified Friedmann equations in \cite{6}. Thus, we insert equation (\ref{Fbi8}) into  equation (\ref{eq29}), and instead of the second term of the above equation (\ref{eq29}), we will have

\begin{flalign}\label{eq30}
 \frac{\ddot{a}}{a}=&\frac{1}{1+F({\cal R} )}\bigg[\frac{-F({\cal R})R}{6}+\frac{F({\cal R}){\cal R}}{3}-\frac{f({\cal R})}{6}-F^{\prime\prime}({\cal R})\dot{{\cal R}}^{2} -F^{\prime}({\cal R})\ddot{{\cal R}}-2HF^{\prime}({\cal R})\dot{{\cal R}}\nonumber\\
&-\kappa^{2}(\frac{P}{2}+\frac{\rho}{6})+\frac{3}{4F({\cal R})}F^{\prime}({\cal R})^{2}\dot{{\cal R} }^{2}\bigg]     .
 \end{flalign}
Using (\ref{Fbi7}) to omit ${\cal R}$ and with some simplification, we can find the following expression

 \begin{equation}\label{Fbi31}
 \frac{\ddot{a}}{a}=\frac{1}{1+F({\cal R})}\bigg[\frac{F({\cal R})R}{6}-\frac{f({\cal R})}{6}+HF^{\prime}({\cal R})\dot{{\cal R}}-
\kappa^{2}\left(\frac{P}{2}+\frac{\rho}{6}\right)+\frac{F^{\prime}({\cal R})^{2}\dot{{\cal R}}^{2}}{4F({\cal R})}\bigg],
  \end{equation}
which is consistent with the final result for the Raychaudhuri equation in $f(R, {\cal R})$ gravity by means of the standard form of the modified Friedmann equations in \cite{6}.


\subsection{GDE for null vector fields}

In this subsection, we calculate the GDE  for past-directed  null vector fields where we have $ V^{\alpha}=k^{\alpha}, k_{\alpha}k^{\alpha}=0 $, so equation (\ref{eq21d}) reduces to

\begin{flalign}\label{eq32}
		R^{\alpha}_{\,\beta\gamma\delta}\kappa^{\beta}\eta^{\gamma}k^{\delta}=&\frac{\eta^{\alpha}E^{2}}{2\left(1+F({\cal R})\right)}\bigg [\kappa^{2}(\rho + P)-\frac{3}{2F({\cal R})}F^{\prime}({\cal R})^{2}\dot{{\cal R}}^{2}+F^{\prime}({\cal R})\ddot{{\cal R}}
		+F^{\prime\prime}({\cal R})\dot{{\cal R}}^{2}-HF^{\prime}({\cal R})\dot{{\cal R}}\bigg].
\end{flalign}
If we consider  $\eta^{\alpha}  $ as $ \eta^{\alpha}=\eta e^{\alpha} , e_{\alpha}e^{\alpha}=1 ,e_{\alpha} u^{\alpha}=e_{\alpha} k^{\alpha} =0$ and $  \frac{D e^{\alpha}}{D\nu}=0 $ equation (\ref{eq22})  reduces to

\begin{flalign}\label{eq33}
\frac{D^{2}\eta^{\alpha}}{D\nu^{2}}=-\frac{\eta^{\alpha}E^{2}}{2\left(1+F({\cal R})\right)}\bigg[\kappa^{2}\left(\rho + P\right)-\frac{3}{2F({\cal R})}F^{\prime}({\cal R})^{2}\dot{{\cal R}}^{2}+F^{\prime}({\cal R})\ddot{{\cal R}}
		+F^{\prime\prime}({\cal R})\dot{{\cal R}}^{2}-HF^{\prime}({\cal R})\dot{{\cal R}}\bigg].
\end{flalign}
In the case of  GR discussed in \cite{7}, all null geodesics experience convergence, provided that $ \kappa(\rho+P)>0 $ and thus the focusing condition for $ f(R,{\cal R}) $ gravity, is

\begin{flalign}\label{eq34}
\frac{1}{\left(1+F({\cal R})\right)}\bigg[\kappa^{2}(\rho + P)-\frac{3}{2F({\cal R})}F^{\prime}({\cal R})^{2}\dot{{\cal R}}^{2}+F^{\prime}({\cal R})\ddot{{\cal R}}+F^{\prime\prime}({\cal R})\dot{{\cal R}}^{2}-HF^{\prime}({\cal R})\dot{{\cal R}}\bigg] >0.
		\end{flalign}
In order to compare with cosmological  observations, we write equation (\ref{eq33}) as a function of the redshift parameter $z$. Differential operators can be used as follows

 \begin{equation}\label{eq35}
\frac{d}{d\nu}=\frac{dz}{d\nu}\frac{d}{dz},
\end{equation}

 \begin{flalign}\label{eq36}
\frac{d^{2}}{d\nu^{2}}&=\frac{dz}{d\nu}\frac{d}{dz}(\frac{d}{d\nu})\nonumber\\
&=(\frac{d\nu}{dz})^{-2}\bigg[-(\frac{d\nu}{dz})^{-1}\frac{d^{2}\nu}{dz^{2}}\frac{d}{dz}+\frac{d^{2}}{dz^{2}}\bigg].
\end{flalign}
For null geodesics we have

\begin{equation}\label{eq37}
(1+z)=\frac{a_{0}}{a}=\frac{E_{0}}{E}\longrightarrow \frac{dz}{1+z}=-\frac{da}{a},
\end{equation}
and we know $ \frac{dt}{d\nu}=E_{0}(1+z) $, so we can get

\begin{equation}\label{eq38}
\frac{d\nu}{dz}=\frac{1}{E_{0}H(1+z)^{2}},
\end{equation}
and
\begin{equation}\label{eq39}
\frac{d^{2}\nu}{dz^{2}}=-\frac{1}{E_{0}H(1+z)^{3}}\bigg[\frac{1}{H}(1+z)\frac{dH}{dz}+2\bigg].
\end{equation}
First, we get $ \frac{dH}{dz} $ and then put it into the equation (\ref{eq39})

\begin{equation}\label{eq40}
\frac{dH}{dz}=- \frac{1}{H(1+z)}\frac{dH}{dt}.
\end{equation}
Defining the Hubble parameter $H=\frac{\dot{a}}{a}$

\begin{equation}\label{eq41}
\dot{H}= \frac{dH}{dt}=\frac{d}{dt}\frac{\dot{a}}{a}=\frac{\ddot{a}}{a}-H^{2}.
\end{equation}
From the equations (\ref{Fbi31}) and (\ref{eq41}), we can write

\begin{flalign}\label{eq42}
\dot{H}=& \frac{1}{1+F({\cal R})}\bigg[\frac{F({\cal R})R}{6}+\frac{F^{\prime}({\cal R})^{2}\dot{{\cal R}}^{2}}{4F({\cal R})}+HF^{\prime}({\cal R})\dot{{\cal R}}-\frac{P}{2}-\frac{\rho}{6}-\frac{f({\cal R})}{6}\bigg]-H^{2}.
\end{flalign}
Finally, equation (\ref{eq39}) is written as follows

\begin{flalign}\label{eq43}
 \frac{d^{2}\nu}{dz^{2}}=&-\frac{1}{E_{0}H(1+z)^{3}}\bigg[\frac{-1}{H^{2}(1+F({\cal R}))}\bigg(\frac{F({\cal R})R}{6}+\frac{F^{\prime}({\cal R})^{2}\dot{{\cal R}}^{2}}{4F({\cal R})}+HF^{\prime}({\cal R})\dot{{\cal R}}-\frac{P}{2}\nonumber\\
& -\frac{\rho}{6}-\frac{f({\cal R})}{6}\bigg)+3\bigg].
\end{flalign}
So
\begin{flalign}\label{eq44}
\frac{d^{2}\eta}{d\nu^{2}}=&(EH(1+z))^{2}\bigg[\frac{-1}{1+z}\bigg(\frac{1}{H^{2}(1+F({\cal R}))}\bigg(\frac{F({\cal R})R}{6}+\frac{F^{\prime}({\cal R})^{2}\dot{{\cal R}}^{2}}{4F({\cal R})} +HF^{\prime}({\cal R})\dot{{\cal R}}\nonumber\\
&-\frac{P}{2}-\frac{\rho}{6}-\frac{f({\cal R})}{6}\bigg)+3\bigg)\frac{d\eta}{dz}+\frac{d^{2}\eta}{dz^{2}}\bigg].
\end{flalign}
According to the calculations, equation (\ref{eq33}) is written as

\begin{flalign}\label{eq45}
&\frac{d^{2}\eta}{dz^{2}}+\frac{3}{1+z}\bigg[1+\frac{1}{3H^{2}(1+F({\cal R}))}\bigg(-\frac{F({\cal R})R}{6}-\frac{F^{\prime}({\cal R} )^{2}\dot{{\cal R}}^{2}}{4F({\cal R})} -HF^{\prime}({\cal R})\dot{{\cal R}}+\kappa^{2}\left(\frac{\rho}{6}+\frac{P}{2}\right)+\frac{f({\cal R})}{6}\bigg)\bigg]\frac{d\eta}{dz}\nonumber\\
&+\frac{\eta}{2H^{2}(1+z)^{2}(1+F({\cal R}))}\bigg[\kappa^{2}(\rho +P)-\frac{3F^{\prime}({\cal R})^{2}\dot{{\cal R}}^{2}}{2F({\cal R})}+ F^{\prime\prime}({\cal R})\dot{{\cal R}}^{2}+F^{\prime}({\cal R})\ddot{{\cal R}}-HF^{\prime}({\cal R})\dot{{\cal R}}\bigg]=0.
\end{flalign}
The expression of energy density and pressure can be considered from \cite{3,4} as

\begin{equation}\label{eq46}
\kappa^{2}\rho =3H_{0}^{2}\left(\Omega_{m_{0}}(1+z)^{3}+\Omega_{r_{0}}(1+z)^{4}\right),        ~~~         \kappa^{2}P =H_{0}^{2}\Omega_{r_{0}}(1+z)^{4},
\end{equation}
and
\begin{equation}\label{eq47}
\Omega_{K_{0}}=-\frac{K}{H_{0}^{2}a_{0}^{2}},
\end{equation}
where $\Omega_{m_{0}}$  and $\Omega_{r_{0}}$ stand  for the dimensionless cosmological density parameters and, the
labels  {\rm m} and {\rm r} refer to the matter and radiation, respectively.  By using the above equations, the null GDE equation (\ref{eq45}) reads

\begin{flalign}\label{eq48}
&\frac{d^{2}\eta}{dz^{2}}+{\cal P}(H,R,z,{\cal R})\frac{d\eta}{dz}+{\cal Q}(H,R,z,{\cal R})\eta =0,
\end{flalign}
with
\begin{flalign}\label{eq49}
{\cal P}(H,R,z,{\cal R})=&\frac{1}{1+z}\bigg[\frac{4\Omega_{r_{0}}(1+z)^{4}+\frac{7}{2}\Omega_{m_{0}}(1+z)^{3}+3(1+F({\cal R}))\Omega_{K_{0}}(1+z)^{2}}{(1+F({\cal R}))\Omega_{K_{0}}(1+z)^{2}+\Omega_{r_{0}}(1+z)^{4}+\Omega_{m_{0}}(1+z)^{3}+\Omega_{DE}}\bigg] \nonumber\\
&+\frac{1}{1+z}\bigg[\frac{2\Omega_{DE}-\frac{3F^{\prime}({\cal R})^{2}\dot{{\cal R}}^{2}}{4F({\cal R})H_{0}^{2}} -\frac{HF^{\prime}({\cal R})\dot{{\cal R}}}{2H_{0}^{2}}+\frac{F^{\prime\prime}({\cal R})\dot{{\cal R}}^{2}}{2H_{0}^{2}}}{(1+F({\cal R}))\Omega_{K_{0}}(1+z)^{2}+\Omega_{r_{0}}(1+z)^{4}+\Omega_{m_{0}}(1+z)^{3}+\Omega_{DE}}\bigg],
\end{flalign}

\begin{flalign}\label{eq50}
{\cal Q}(H,R,z,{\cal R})=&\frac{1}{2(1+z)^{2}}\bigg[\frac{4\Omega_{r_{0}}(1+z)^{4}+3\Omega_{m_{0}}(1+z)^{3}}{(1+F({\cal R}))\Omega_{K_{0}}(1+z)^{2}+\Omega_{r_{0}}(1+z)^{4}+\Omega_{m_{0}}(1+z)^{3}+\Omega_{DE}}\bigg]\nonumber\\
&+\frac{1}{2(1+z)^{2}}\bigg[\frac{2\Omega_{DE}-\frac{HF^{\prime}({\cal R})\dot{{\cal R}}}{H_{0}^{2}}+\frac{f({\cal R})}{3H_{0}^{2}}-\frac{F({\cal R})R}{3H_{0}^{2}}+\frac{F^{\prime}({\cal R})^{2}\dot{{\cal R}}^{2}}{F({\cal R})H_{0}^{2}}}{(1+F({\cal R}))\Omega_{K_{0}}(1+z)^{2}+\Omega_{r_{0}}(1+z)^{4}+\Omega_{m_{0}}(1+z)^{3}+\Omega_{DE}}\bigg],
\end{flalign}
in which

\begin{flalign}\label{eq51}
\Omega_{DE}=&\frac{1}{H_{0}^{2}}\bigg[\frac{F({\cal R} )R}{6}-\frac{f({\cal R})}{6}-\frac{F^{\prime}({\cal R})^{2}\dot{{\cal R}}^{2}}{4F({\cal R})}+\frac{1}{2}\bigg(F^{\prime\prime}({\cal R})\dot{{\cal R}}^{2}+F^{\prime}({\cal R})\ddot{{\cal R}}+HF^{\prime}({\cal R})\dot{{\cal R}} \bigg)\bigg].
\end{flalign}
From \cite{6}, we redefine

\begin{equation}\label{eq52}
{\cal H}=H+\frac{F^{\prime}({\cal R})\dot{{\cal R}}}{2F({\cal R})}.
\end{equation}
As a result, the modified first Friedmann equation is obtained as

\begin{equation}\label{eq53}
H^{2}=-\frac{K}{a^{2}}(1+F({\cal R}))-F({\cal R}){\cal H}^{2}+\frac{1}{6}(F({\cal R}){\cal R} -f({\cal R}))+\frac{\kappa^{2}\rho}{3}.
\end{equation}
As before, contracting ${\cal R}_{\mu\nu}$ with  $ g^{\mu\nu} $ in (\ref{Fbi7}) leads to the following relation  \cite{4k}

\begin{equation}\label{eq54}
{\cal R}=R-\frac{3F^{\prime}({\cal R})^{2}\dot{{\cal R}}^{2}}{2F({\cal R})^{2}}+\frac{3}{F({\cal R})}\bigg(F^{\prime\prime}({\cal R})\dot{{\cal R}}^{2}+F^{\prime}({\cal R})\ddot{{\cal R}}+3HF^{\prime}({\cal R})\dot{{\cal R}}\bigg).
\end{equation}
From (\ref{eq51})-(\ref{eq54}), it is possible to obtain the first modified Friedmann equation in $R+f\left({\cal R}\right)$  as

\begin{flalign}\label{eq55}
H^{2}=&\frac{H_{0}^{2}}{1+F({\cal R})}\bigg[\Omega_{K_{0}}(1+z)^{2}\left(1+F({\cal R})\right)+\Omega_{m_{0}}(1+z)^{3}+\Omega_{r_{0}}(1+z)^{4}+\Omega_{DE}\bigg].
\end{flalign}
As a particular model, let us consider the case $ f({\cal R})={\cal R}-2\Lambda $, whereby $\Lambda$ is cosmological constant parameter that $\Omega_{\Lambda}=\frac{\Lambda}{3H_{0}^{2}}$. Therefore, we have

\begin{equation}\label{eq56}
F({\cal R})=1~~~~~~~~F^{\prime}({\cal R})=0,
\end{equation}
so
\begin{flalign}\label{eq57}
\Omega_{DE}=\frac{1}{6H_{0}^{2}}\left[R-\left({\cal R}-2\Lambda\right)\right].
\end{flalign}
Equations (\ref{eq49}) and (\ref{eq50}) are reduced as follows

\begin{flalign}\label{eq58}
&{\cal P}(H,R,z,{\cal R})=\frac{1}{1+z}\bigg[\frac{4\Omega_{r_{0}}(1+z)^{4}+\frac{7}{2}\Omega_{m_{0}}(1+z)^{3}+6\Omega_{K_{0}}(1+z)^{2}+
2\Omega_{DE}}{2\Omega_{K_{0}}(1+z)^{2}+\Omega_{r_{0}}(1+z)^{4}+\Omega_{m_{0}}(1+z)^{3}+\Omega_{DE}}\bigg]  ,
\end{flalign}

\begin{flalign}\label{eq59}
&{\cal Q}(H,R,z,{\cal R})=\frac{1}{\left(1+z\right)^{2}}\left[\frac{2\Omega_{r_{0}}(1+z)^{2}+\frac{3}{2}\Omega_{m_{0}}(1+z)^{3}
+\frac{\Omega_{DE}}{2}-\frac{\Omega_{\Lambda}}{2}}{\Omega_{r_{0}}(1+z)^{2}\left(z^{2}+2z-1\right)+\Omega_{m_{0}}(1+z)^{2} (z-1)+2(1+z)^{2}+\Omega_{DE}}\right].
\end{flalign}
For the particular choices $ \Omega_{\Lambda}=0 $ and $ \Omega_{K_{0}}=1-  \Omega_{m_{0}}- \Omega_{r_{0}} $ we can find the modified {\it Mattig} relation. Thus,

\begin{flalign}\label{eq60}
&{\cal \tilde{P}}(z)=\frac{1}{1+z}\bigg[\frac{4\Omega_{r_{0}}(1+z)^{2}\left(z^{2}+2z-\frac{1}{2}\right)+\frac{7z-5}{2}\Omega_{m_{0}}(1+z)^{2}+
6(1+z)^{2}+2\Omega_{DE}}{2(1+z)^{2}+
\Omega_{r_{0}}(1+z)^{2}\left(z^{2}+2z-1\right)+\Omega_{m_{0}}(1+z)^{2}(z-1)+\Omega_{DE}}\bigg],
\end{flalign}

\begin{flalign}\label{eq61}
&{\cal \tilde{Q}}(z)=\frac{4\Omega_{r_{0}}(1+z)^{2}+3\Omega_{m_{0}}(1+z)^{3}+\Omega_{DE}}{2(1+z)^{2}\bigg[2(1+z)^{2}+
\Omega_{r_{0}}(1+z)^{2}(z^{2}+2z-1)+\Omega_{m_{0}}\left(1+z\right)^{2} \left(z-1\right)+\Omega_{DE}\bigg]}.
\end{flalign}
Finally, we find the modified {\it Mattig} relation in $f\left(R,{\cal R}\right)$ as follows

\begin{flalign}\label{eq62}
\frac{d^{2}\eta}{dz^{2}}+{\cal \tilde{P}}(z) \frac{d\eta}{dz}
+{\cal \tilde{Q}}(z)\eta =0.
\end{flalign}
It is worth noting to note that, for a spherically symmetric space-time similar to the FLRW universe, the deviation vector magnitude $\eta$ is proportional to the proper area $dA$ of a source with a redshift $z$ as $d\eta \propto \sqrt{dA}$ which leads to the definition of the observer area-distance $ r_0(z)$ with the expression

\begin{flalign}\label{eq63}
r_0(z)=\sqrt{\left|\frac{dA_0(z)}{d\Omega}\right|}=\left|\frac{\eta(z')|_{z'=z}}{d\eta(z')/dl|_{z'=0}}\right|  ,
\end{flalign}
where $A_0$  represents the area of the object, and $\Omega_0$ is the solid angle \cite{7, 111}. Thus, by implying the relation
$|d/dl|=E^{-1}_0 (1+ z)^{-1}d/d\nu=H (1 + z)d/dz$, where in $dl= a(t)dr$, while assuming that the deviation vector to be zero at $z=0$, equation (\ref{eq63}) can be written as follows
\begin{flalign}\label{eq64}
r_0(z)=\left|\frac{\eta(z)}{H_0d\eta(z')/dz'|_{z'=0}}\right|.
\end{flalign}
This equation denotes the observed area-distance $r_0(z)$ as a function of $z$ in units of the present-day Hubble radius.



\subsection{Numerical solutions of the GDE for $f(R,{\cal R})$ gravity}

Clearly, to find the solutions of (\ref{eq48}) (the null GDE), we are supposed to consider $f\left(R,{\cal R}\right)$ forms. The standard  form is the case $f\left(R,{\cal R}\right)=R-2\Lambda$. In this case, we obtain the trivial solution, {\it i.e.}, the $\Lambda$CDM model. Another functional form  is the case $f\left(R,{\cal R}\right)=f(R)$, which was considered in \cite{4m}. In order to discover the new properties of $f\left(R,{\cal R}\right)$ gravity, we should consider the cases with ${\cal R}\neq 0$.\\
In order to examine our study, we consider the numerical solutions of the GDE by taking the hybrid Metric-Palatini function as $f\left(R,{\cal R}\right)=R+{\cal R}$, thus, equation (\ref{eq48}) is reduced to

\begin{flalign}\label{eq67}
&\frac{d^{2}\eta}{dz^{2}}+{\cal P}_{0}(H,R,z,{\cal R})\frac{d\eta}{dz}+{\cal Q}_{0}(H,R,z,{\cal R})\eta =0,
\end{flalign}
where we define

\begin{flalign}\label{eq68}
{\cal P}_{0}(H,R,z,{\cal R})=&\frac{1}{1+z}\bigg[\frac{4\Omega_{r_{0}}(1+z)^{4}+\frac{7}{2}\Omega_{m_{0}}(1+z)^{3}+ 6\Omega_{K_{0}}(1+z)^{2}+2\Omega_{DE}}{ 2\Omega_{K_{0}}(1+z)^{2}+\Omega_{r_{0}}(1+z)^{4}+\Omega_{m_{0}}(1+z)^{3}+\Omega_{DE}}\bigg],
\end{flalign}
and
\begin{flalign}\label{eq69}
{\cal Q}_{0}\left(H,R,z,{\cal R}\right)=&\frac{1}{\left(1+z\right)^{2}}\bigg[\frac{2\Omega_{r_{0}}(1+z)^{4}+\frac{3}{2}\Omega_{m_{0}}(1+z)^{3}+2\Omega_{DE}-\Omega_{\Lambda}}{ 2\Omega_{K_{0}}(1+z)^{2}+\Omega_{r_{0}}(1+z)^{4}+\Omega_{m_{0}}(1+z)^{3}+\Omega_{DE}}\bigg].
\end{flalign}
Now we can solve the equation (\ref{eq67}) numerically to find the evolution of $\eta(z)$
and $r_{0}(z)$ as functions of $z$ and the result is plotted in figure $1$.

\begin{figure*}[ht]
  \centering
  \includegraphics[width=2in]{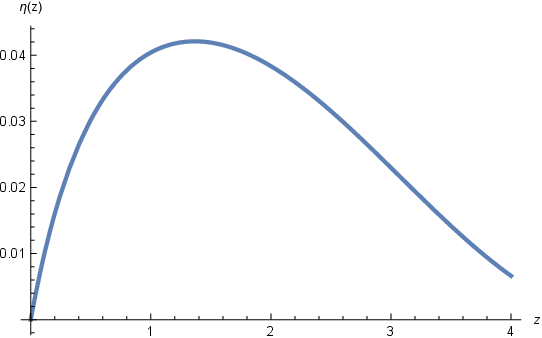}\hspace{1.9cm}
  \includegraphics[width=2in]{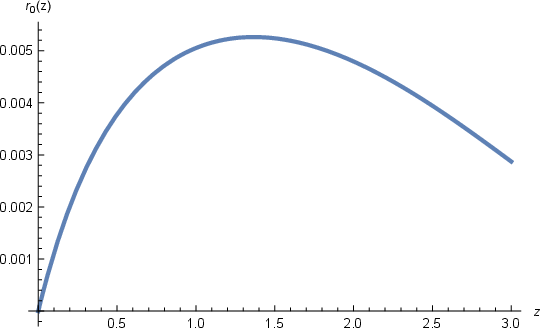}
    \caption{The Deviation vector $\eta(z)$ evolution plot (left) and the observer area-distance $r_{0}(z)$ evolution plot (right).
    With the numerical value consideration $\Omega_{m_{0}}=0.3$, $\Omega_{r_{0}}=0$, $\Omega_{\Lambda}=0$, $\Omega_{K_{0}}=0.001$, $\Omega_{DE}=0.7$ and $\frac{d\eta(z)}{dz}|_{z=0}=0.1$.}
  \label{stable}
\end{figure*}

\section{Scalar-tensor representation of $ f(R,{\cal R}) $ gravity}

We start from the following action \cite{8}, in which we have a general function with two variables, metric and Palatini curvature scalars. In this section, we take a look at how this generalization can be considered dynamically equivalent to a gravitational theory with two scalar fields. According to \cite{8}, the general form of Metric-Palatini action can be written as follows

\begin{equation}\label{eq70}
S=\frac{1}{2\kappa^{2}}\int d^4{x}\sqrt{-g}f\left(R,{\cal R}\right)+S_{M}.
\end{equation}
The variation of the action (\ref{eq70}) with respect to the metric and connection, the field equations can be respectively written as

\begin{flalign}\label{eq75}
\frac{\partial f}{\partial R}R_{\mu\nu}-\frac{1}{2}g_{\mu\nu}f-\left(\nabla_{\mu}\nabla_{\nu}-g_{\mu\nu}\square\right)\frac{\partial f}{\partial R}+\frac{\partial f}{\partial {\cal R}}{\cal R}_{\mu\nu}=\kappa^{2} T_{\mu\nu},
\end{flalign}

\begin{equation}\label{eq76}
\hat{\nabla}_{\lambda}\bigg(\sqrt{-g}\frac{\partial f}{\partial{\cal R}}g^{\mu\nu}\bigg)=0,
\end{equation}
where the covariant derivative $\hat{\nabla}$ is related to the metric $ h_{\mu\nu}=\frac{\partial f}{\partial{\cal R}}g_{\mu\nu} $.
We can take the action with two scalar fields $\alpha $ and  $\beta$ as follows

\begin{equation}\label{eq77}
S=\frac{1}{2\kappa^{2}}\int d^4{x}\sqrt{-g}\bigg[f\left(\alpha,\beta\right)+\frac{\partial f\left(\alpha,\beta\right)}{\partial \alpha}\left(R-\alpha\right)+\frac{\partial f\left(\alpha,\beta\right)}{\partial \beta}\left({\cal R}-\beta\right)\bigg]+S_{M}.
\end{equation}
Then, it is possible to obtain the field equations by variation with respect to $ \alpha $ and $  \beta$ from the action of (\ref{eq77}). We define the two new scalar fields as
\begin{equation}\label{eq78}
\chi =\frac{\partial f(\alpha,\beta)}{\partial \alpha} \hspace{4mm} \textrm{and} \hspace{4mm} \xi =-\frac{\partial f(\alpha,\beta)}{\partial \beta}.
\end{equation}
According to \cite{8}, the action (\ref{eq77}) can be rewritten as

\begin{equation}\label{eq79}
S=\frac{1}{2\kappa^{2}}\int d^4{x}\sqrt{-g}\bigg[\left(\chi -\xi\right)R-\frac{3}{2\xi}(\partial\xi)^{2}-V(\chi,\xi)\bigg]+S_{M},
\end{equation}
given that $ V(\chi,\xi) $ is considered as

\begin{equation}\label{eq80}
V(\chi,\xi)=-f\left(\alpha(\chi),\beta(\xi)\right)+\chi\alpha(\chi)-\xi\beta(\xi).
\end{equation}
We define a new scalar field as $ \phi =\chi-\xi $ and we can perform a conformal transformation to exchange from Jordan's framework to Einstein's as follows

\begin{equation}\label{eq81}
g_{\mu\nu}\rightarrow \tilde{g}_{\mu\nu}=\phi  g_{\mu\nu}.
\end{equation}
Therefore, we have

\begin{equation}\label{eq82}
S=\frac{1}{2\kappa^{2}}\int d^4{x}\sqrt{-\tilde{g}}\bigg[\tilde{R}-\frac{3}{2\phi^{2}}(\partial\phi)^{2}-\frac{3}{2\phi\xi}(\partial\xi)^{2}-
\frac{W(\phi,\xi)}{\phi^{2}}\bigg]+S_{M}.
\end{equation}
Now, we redefine two new scalar fields as

\begin{equation}\label{eq83}
\tilde{\phi}=\sqrt{\frac{3}{2}}\frac{\ln\phi}{k}\hspace{4mm}\textrm{and}\hspace{4mm}\tilde{\xi}=\frac{2\sqrt{2}}{k}\sqrt{\xi}.
\end{equation}
Finally, we have

\begin{equation}\label{eq84}
S=\int d^4{x}\sqrt{-\tilde{g}}\bigg[\frac{1}{2\kappa^{2}}\tilde{R}-\frac{1}{2}(\tilde\nabla\tilde{\phi})^{2}-
\frac{1}{2}e^{\frac{-\sqrt{2}\kappa\tilde{\phi}}{\sqrt{3}}}(\tilde\nabla\tilde{\xi})^{2}-\tilde{W}(\tilde{\phi},\tilde{\xi})\bigg]+\tilde{S}_{M},
\end{equation}

\begin{equation}\label{eq85}
\tilde{S}_{M}=e^{\frac{-\sqrt{2}\kappa\tilde{\phi}}{3}}S_{M},
\end{equation}
where

\begin{equation}\label{eq86}
\tilde{W}(\tilde{\phi},\tilde{\xi})=
\frac{1}{2\kappa^{2}}e^{-\frac{\sqrt{2}\kappa\tilde{\phi}}{\sqrt{3}}}W\left(e^{\frac{\sqrt{2}\kappa\tilde{\phi}}{\sqrt{3}}},\kappa^{2}\frac{\tilde{\xi}^{2}}{8}\right),
\end{equation}
which is the new potential.

It should be noted that the Brans-Dicke context introduces the scalar field $\phi$ as the Brans-Dicke field and $\xi$ as the inflation. In fact, the action (\ref{eq79}) is usually more extended than the form we considered, which means that it includes a kinetic term for $\phi$ or a more general coupling term between $\phi$ and $R$ \cite{9,10}.
For the sake of simplicity, from now on we will omit the tildes in action (\ref{eq84}). The field equations can be obtained by varying action (\ref{eq84}) with respect to $ g_{\mu\nu} $ as

\begin{equation}\label{eq87}
G_{\mu\nu}=\kappa^{2}\bigg(T^{(\phi)}_{\mu\nu}+e^{-\frac{\sqrt{2}\kappa{\phi}}{\sqrt{3}}}\left(T^{(\xi)}_{\mu\nu}+T_{\mu\nu}\right)- g_{\mu\nu}W
\bigg),
\end{equation}
by considering

\begin{equation}\label{eq88}
T^{(\phi)}_{\mu\nu}=\triangledown_{\mu}\phi \triangledown _{\nu}\phi -\frac{1}{2}g_{\mu\nu}(\nabla\phi)^{2},
\end{equation}

\begin{equation}\label{eq89}
T^{(\xi)}_{\mu\nu}=\triangledown_{\mu}\xi \triangledown _{\nu}\xi -\frac{1}{2}g_{\mu\nu}(\nabla\xi)^{2}.
\end{equation}
In order to extract the geodesic deviation equation (GDE), we first need to calculate $ R_{\mu\nu} $ from the modified Einstein equation (\ref{eq87}).

\begin{equation}\label{eq90}
R_{\mu\nu}=\kappa^{2}T^{(\textrm{tot})}_{\mu\nu}+\frac{1}{2}g_{\mu\nu}\left(R-2\kappa^{2}W\right),
\end{equation}
and
\begin{equation}\label{eq91}
R=\kappa^{2}\left(4W-T^{(\textrm{tot})}\right),
\end{equation}
while
\begin{equation}\label{eq92}
T^{(\textrm{tot})}_{\mu\nu}=T^{(\phi)}_{\mu\nu}+e^{-\frac{\sqrt{2}\kappa{\phi}}{\sqrt{3}}}\left(T^{(\xi)}_{\mu\nu}+T_{\mu\nu}\right).
\end{equation}
To complete our investigation to obtain the GDE in the context of the scalar-tensor theory of $f\left(R,{\cal R}\right)$ gravity, we should carry out the extraction of the product of the Riemann tensor contraction with respect to the normalized tangent vectors and the geodesic deviation vector in this modified theory.

\subsection{GDE in the context of Scalar-tensor theory of $ f(R,{\cal R}) $}

To continue, we will investigate the GDE for the action (\ref{eq84}). First, we calculate the Riemann tensor through equations (\ref{Fbi10}), (\ref{eq90}) and (\ref{eq91}), written in the following form

\begin{equation}\label{eq93}
R^{\alpha}_{\,\beta\gamma\delta}=\frac{\kappa^{2}}{2}\bigg[\delta_{\gamma}^{\alpha}
T_{\delta\beta}^{(\textrm{tot})}-\delta_{\delta}^{\alpha}T_{\gamma\beta}^{(\textrm{tot})}+
g_{\beta\delta}T_{\gamma}^{(\textrm{tot})\alpha}-g_{\beta\gamma}
T_{\delta}^{(\textrm{tot})\alpha}\bigg]+\frac{\kappa^{2}}{3}\bigg[\left(\delta_{\gamma}^{\alpha}g_{\delta\beta}-\delta_{\delta}^{\alpha}
g_{\gamma\beta}\right)\left(W-T^{(\textrm{tot})}\right)\bigg].
\end{equation}
Contracting the Riemann tensor  with the $V^{\beta}\eta^{\gamma}V^{\delta}$ term, the following result is obtained.

\begin{flalign}\label{eq94}
R^{\alpha}_{\,\beta\gamma\delta}V^{\beta}\eta^{\gamma}V^{\delta}= \kappa^{2}\bigg[E^{2}\left(\frac{\dot{\phi}^{2}}{2}+\frac{e^{-\frac{\sqrt{2}\kappa{\phi}}{\sqrt{3}}}}{2}\left(\dot{\xi}^{2}+\rho\ +P\right)\right)+
\epsilon\left(\frac{\dot{\phi}^{2}}{2}+e^{-\frac{\sqrt{2}\kappa{\phi}}{\sqrt{3}}}\left(\frac{\dot{\xi}^{2}}{2}+p\right)
+\frac{\left(W-T^{(\textrm{tot})}\right)}{3}\right)\bigg]\eta^{\alpha}.
\end{flalign}
Again, it is useful to define $\tilde{\rho}_{\textrm{eff}}$ and $\tilde{P}_{\textrm{eff}}$ to reduce the GDE in a comprehensive form. As a result, we can write

\begin{flalign}\label{eq95}
\tilde{\rho}_{\textrm{eff}}=\frac{\dot{\phi}^{2}}{2}+e^{-\frac{\sqrt{2}\kappa{\phi}}{\sqrt{3}}}\left(\frac{\dot{\xi}^{2}}{2}+\rho\right)-
\frac{\left(W-T^{(\textrm{tot})}\right)}{3},
\end{flalign}

\begin{flalign}\label{eq96}
\tilde{P}_{\textrm{eff}}=\frac{\dot{\phi}^{2}}{2}+e^{-\frac{\sqrt{2}k{\phi}}{\sqrt{3}}}\left(\frac{\dot{\xi}^{2}}{2}+P\right)+
\frac{\left(W-T^{(\textrm{tot})}\right)}{3}.
\end{flalign}
Hence the modified \textit{Pirani} equation is obtained as

\begin{flalign}\label{eq97}
R^{\alpha}_{\,\beta\gamma\delta}V^{\beta}\eta^{\gamma}V^{\delta}=
\frac{\kappa^{2}}{2}\left[E^{2}\left(\tilde{\rho}_{\textrm{eff}}+\tilde{P}_{\textrm{eff}}\right)
+\epsilon \left(2\tilde{P}_{\textrm{eff}}\right)\right]\eta^{\alpha},
\end{flalign}
by which the GDE becomes

\begin{flalign}\label{eq98}
\frac{D^{2}\eta}{D\nu^{2}}=-\frac{\kappa^{2}}{2}\left[E^{2}\left(\tilde{\rho}_{\textrm{eff}}+\tilde{P}_{\textrm{eff}}\right)
+\epsilon \left(2\tilde{P}_{\textrm{eff}}\right)\right]\eta^{\alpha}.
\end{flalign}
As before, in the next step we are supposed to find the GDE for fundamental observers with the condition $E^{2}=1$ and $\epsilon=-1$.

\subsection{GDE for fundamental observers}
Now, we are going to find the GDE for fundamental observers in the scalar-tensor theory of $f\left(R,{\cal R}\right)$ gravity by exerting the condition $E^{2}=1$ and $\epsilon=-1$. Subsequently, in this case

\begin{flalign}\label{eq99}
R^{\alpha}_{\,\beta\gamma\delta}u^{\beta}\eta^{\gamma}u^{\delta} =\frac{\kappa^{2}}{2}\left[\tilde{\rho}_{\textrm{eff}}-\tilde{P}_{\textrm{eff}}\right]\eta^{\alpha}.
\end{flalign}
According to the method mentioned in section 2, we will have

\begin{flalign}\label{eq100}
 \frac{\ddot{a}}{a}=\frac{\kappa^{2}}{2}\left[\tilde{P}_{\textrm{eff}}-\tilde{\rho}_{\textrm{eff}}\right].
\end{flalign}
From equation  (\ref{eq92}) we can calculate $T^{(\textrm{tot})} $ as

\begin{equation}\label{eq101}
T^{(\textrm{tot})}=T^{(\phi)}+e^{-\frac{\sqrt{2}\kappa{\phi}}{\sqrt{3}}}\left(T^{(\xi)}+T\right).
\end{equation}
Thus, the above equation can be rewritten as

\begin{equation}\label{eq102}
T^{(\textrm{tot})}=\dot{\phi}^{2}+e^{-\frac{\sqrt{2}\kappa{\phi}}{\sqrt{3}}}\left(\dot{\xi}^{2}+3p-\rho\right).
\end{equation}
Hence, the modified Raychaudhuri equation (\ref{eq100}) can be written as

\begin{flalign}\label{eq103}
 \frac{\ddot{a}}{a}=\kappa^{2}\bigg[\frac{W}{3}-\frac{\dot{\phi}^{2}}{3}-e^{-\frac{\sqrt{2}\kappa{\phi}}{\sqrt{3}}}\left(\frac{\dot{\xi}^{2}}{3}+
 \frac{\rho}{6}+\frac{p}{2}\right)\bigg].
\end{flalign}
Moreover, we are going to check the correctness of the result (\ref{eq103}) using the first and second modified Friedmann equations described in \cite{8}. The cosmological equations are in the following order

\begin{flalign}\label{eq104}
\frac{K}{a^{2}}+H^{2}=\frac{\kappa^{2}}{3}\left(e^{-\frac{\sqrt{2}\kappa{\phi}}{\sqrt{3}}}\left(\frac{\dot{\xi}^{2}}{2}-\rho\right)+
\frac{\dot{\phi}^{2}}{2}+W\right),
\end{flalign}

\begin{flalign}\label{eq105}
\frac{K}{a^{2}}+2\dot{H}+3H^{2}=-\kappa^{2}\left(e^{-\frac{\sqrt{2}\kappa{\phi}}{\sqrt{3}}}\left(\frac{\dot{\xi}^{2}}{2}+p\right)+
\frac{\dot{\phi}^{2}}{2}-W\right),
\end{flalign}
On the other hand, we know that $\frac{K}{a^{2}}=\frac{R}{6}-H^{2}-\frac{\ddot{a}}{a}$ and $\frac{\ddot{a}}{a}=\dot{H}+H^{2}$. Beside the above two cosmological equations (\ref{eq104}) and (\ref{eq105}), we can derive the relation in equation (\ref{eq103}), which shows that the result (\ref{eq103}) is correct.

\subsection{GDE for null vector fields}

In the following, we calculate the GDE for null vector fields in the Scalar-tensor theory of $f\left(R,{\cal R}\right)$ gravity. As in section 2, we have $ V^{\alpha}=k^{\alpha} $ which means that $\epsilon=0$ therefore, the equation (\ref{eq97}) reduces to

\begin{flalign}\label{eq106}
R^{\alpha}_{\,\beta\gamma\delta}k^{\beta}\eta^{\gamma}k^{\delta} =\frac{\kappa^{2}E^{2}}{2}\left[\tilde{\rho}_{\textrm{eff}}+\tilde{P}_{\textrm{eff}}\right]\eta^{\alpha},
\end{flalign}
resulting in

\begin{flalign}\label{eq107}
\frac{d^{2}\eta}{d\nu^{2}}=-\frac{\kappa^{2}E^{2}}{2}\left[\tilde{\rho}_{\textrm{eff}}+\tilde{P}_{\textrm{eff}}\right]\eta^{\alpha}.
\end{flalign}
As well, in this case the past-directed null geodesics experience focusing if the null energy condition is satisfied as

\begin{flalign}\label{eq108}
\tilde{\rho}_{\textrm{eff}}+\tilde{P}_{\textrm{eff}}>0.
\end{flalign}
Equivalently, we can say

\begin{flalign}\label{eq109}
\rho+P>-e^{\frac{\sqrt{2}\kappa{\phi}}{\sqrt{3}}}\dot{\phi}^{2}+\dot{\xi}^{2}.
\end{flalign}
Here, similar to the approach that it represented in section 2, we obtain the GDE for null vector fields in the framework of the scalar-tensor representation of the model.
To continue, we first obtain $\dot{H}$ in this model as
\begin{flalign}\label{eq110}
\dot{H}=\kappa^{2}\bigg[\frac{e^{\frac{-\sqrt{2}\kappa\phi}{\sqrt{3}}}\left(p-\rho\right)}{2}+\frac{W-T^{\textrm{tot}}}{3}\bigg]-H^{2}.
\end{flalign}
Then we have
\begin{flalign}\label{eq111}
\frac{d^{2}\nu}{dz^{2}}=\frac{-3}{EH\left(1+z\right)^{2}}\left[1+\frac{\kappa^{2}}{3H^{2}}\left(\frac{e^{\frac{\sqrt{2}\kappa\phi}{\sqrt{3}}}\left(
\rho-P\right)}{2}+\frac{T^{(\textrm{tot})}-W}{3}\right)\right].
\end{flalign}
Finally, by using the equations (\ref{eq39}) and (\ref{eq38}), $ \frac{d^{2}\eta}{d\nu^{2}} $ can be obtained as follows

\begin{flalign}\label{eq112}
\frac{d^{2}\eta}{d\nu^{2}}=\left(EH(1+z)\right)^{2}\bigg[\frac{3}{1+z}\bigg(1-\frac{\kappa^{2}}{3H^{2}}
\left(\frac{e^{-\frac{\sqrt{2}\kappa{\phi}}{\sqrt{3}}}(p-\rho)
}{2}+\frac{W-T^{\textrm{tot}}}{3}\right)\bigg)\frac{d\eta}{dz}+\frac{d^{2}\eta}{dz^{2}}\bigg].
\end{flalign}
Using (\ref{eq107}) we can imply

\begin{flalign}\label{eq113}
\frac{d^{2}\eta}{dz^{2}}+\frac{3}{1+z}\bigg[1-\frac{\kappa^{2}}{6H^{2}}\bigg(\tilde{P}_{\textrm{eff}}-
\tilde{\rho}_{\textrm{eff}}\bigg)\bigg]\frac{d\eta}{dz}+\frac{\kappa^{2}}{2H^{2}(1+z)^{2}}\left[\tilde{\rho}_{\textrm{eff}}+
\tilde{P}_{\textrm{eff}}\right]\eta =0,
\end{flalign}

\begin{flalign}\label{eq114}
&\frac{d^{2}\eta}{dz^{2}}+{\cal P}(H,R,z)\frac{d\eta}{dz}+{\cal Q}(H,R,z,)\eta =0,
\end{flalign}
with
\begin{flalign}\label{eq115}
{\cal P}(H,R,z)=\frac{1}{1+z}\bigg[\frac{\left(-2\Omega_{r_{0}}(1+z)^{4}
-\frac{5}{2}\Omega_{m_{0}}(1+z)^{3}\right)e^{-\frac{\sqrt{2}\kappa{\phi}}{\sqrt{3}}}+
3\Omega_{K_{0}}(1+z)^{2}+5\Omega_{DE}-\frac{\kappa^{2}W}{H^{2}_{0}}}{-e^{-\frac{\sqrt{2}\kappa{\phi}}{\sqrt{3}}}\left(\Omega_{r_{0}}(1+z)^{4}+
\Omega_{m_{0}}(1+z)^{3}\right)+\Omega_{K_{0}}(1+z)^{2}+\Omega_{DE}}\bigg],
\end{flalign}

\begin{flalign}\label{eq116}
{\cal Q}(H,R,z,)=\frac{1}{\left(1+z\right)^{2}}\left[\frac{e^{\frac{-\sqrt{2}\kappa\phi}{\sqrt{3}}}\left(\frac{3}{4}\Omega_{m_{0}}\left(1+z\right)^{3}+
\Omega_{r_{0}}\left(1+z\right)^{4}\right)+\frac{3}{2}\Omega_{DE}-\frac{\kappa^{2}W}{2H_{0}^{2}}}{-e^{-\frac{\sqrt{2}\kappa{\phi}}{\sqrt{3}}}
\left(\Omega_{r_{0}}(1+z)^{4}+
\Omega_{m_{0}}(1+z)^{3}\right)+\Omega_{K_{0}}(1+z)^{2}+\Omega_{DE}}\right],
\end{flalign}
and $ H^{2} $ given in the equation (\ref{eq104}) is rewritten as

\begin{flalign}\label{eq117}
 H^{2}=H^{2}_{0}\bigg[-e^{-\frac{\sqrt{2}\kappa{\phi}}{\sqrt{3}}}\left(\Omega_{r_{0}}(1+z)^{4}+\Omega_{m_{0}}(1+z)^{3}\right)+
 \Omega_{K_{0}}(1+z)^{2}+\Omega_{DE}\bigg],
\end{flalign}
where

\begin{flalign}\label{eq118}
\Omega_{DE}=\frac{\kappa^{2}}{H_{0}^{2}}\bigg[\dot{\phi}^{2}+2W+e^{-\frac{\sqrt{2}\kappa{\phi}}{\sqrt{3}}}\dot{\xi}^{2}\bigg].
\end{flalign}
For the generalized case $\Omega_{DE}\neq 0$, $\Omega(m_{0})\neq 0$, $\Omega(K_{0})\neq 0$, and $\Omega(r_{0})\neq 0$ we can investigate the solution for the equation (\ref{eq114}) by applying numerical analysis similar to section 2. Thus we have plotted the deviation vector and the observer are-distance evolution in terms of the redshift $z$ depicted in figure 2.

\begin{figure*}[ht]
  \centering
  \includegraphics[width=2in]{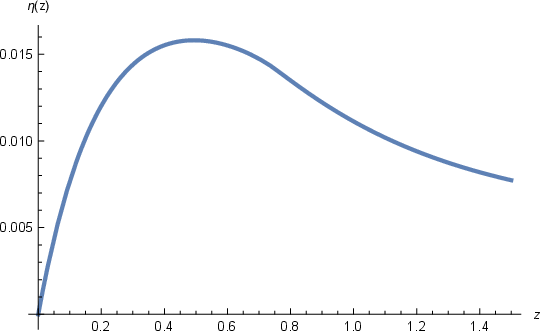}\hspace{1.9cm}
  \includegraphics[width=2in]{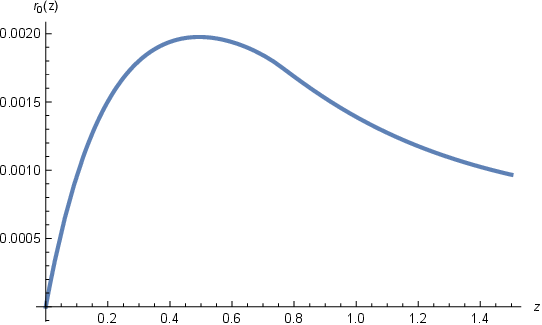}
    \caption{The Deviation vector $\eta(z)$ evolution plot (left) and the observer area-distance $r_{0}(z)$ evolution plot (right).
   With the numerical value consideration $\Omega_{m_{0}}=0.3$, $\Omega_{r_{0}}=0$, $\Omega_{\Lambda}=0$, $\Omega_{K_{0}}=0.001$, $\Omega_{DE}=0.7$ and $\frac{d\eta(z)}{dz}|_{z=0}=0.1$ with the assumption $\kappa\phi_0=\kappa^{2}W=1$.}
  \label{stable}
\end{figure*}

\section{Conclusions}

In this paper, we have investigated the GDE as a basic equation in hybrid Metric-Palatini gravity $f\left(R,\cal{R}\right)$ and the scalar-tensor representation of $f\left(R,\cal{R}\right)$ gravity in order to study the relationship between the Riemann curvature tensor and the relative acceleration between two nearby test particles. First, we studied the field equations in $f\left(R,\cal{R}\right)$ gravity considering an action including a general function $f(\cal{R})$ besides the Einstein-Hilbert action one in the form of $R+f(\cal{R})$. We then obtained the GDE generalized expression for $R+f(\cal{R})$ in the context of the FLRW universe with perfect fluid energy-momentum tensor in which the effective energy density and pressure are given by $\rho_{\textrm{eff}}$ and $P_{\textrm{eff}}$, respectively. In the next part, we set $f\left(R,\cal{R}\right)=R-2\Lambda$ and checked the correctness of our result by analogy with the GR scenario.
In subsection $2.2$, we found the generalized GDE for fundamental observers besides the modified Pirani equation and the Raychaudhuri equation. We also studied the GDE for null vector fields to extract the null GDE equation and we have investigated the focusing condition for this model in which the geodesics experience convergence in addition to the modified Mattig relation.
Moreover, we proposed a particular case $f(R,{\cal R})=R+{\cal R}$  in order to study the numerical behavior of the deviation vector $\eta(z)$ and the observer area-distance $r_{0}(z)$ as a function of redshift, and the result is plotted in figure 1. The appearance of a peak point for $\eta(z)$ and $r_{0}(z)$ in a specified redshift implies that there existed maximum values for $\eta(z)$ and $r_{0}(z)$ in the past when our universe was experiencing the inflation regime. After this, the universe exited the inflation regime, and the deviation vector gradually decreased by increasing $z$. In section 3 we reviewed a dynamically equivalent approach to $f\left(R,\cal{R}\right)$ gravity with two scalar fields, and we found the GDE in the scalar-tensor representation of the model. We have repeated the main approach of this study for this modification and the deviation vector  $\eta(z)$ and the observer area-distance $r_{0}(z)$ versus redshift are plotted in figure 2.
\\
To summarize our results, in this work, we have studied the observed area-distance of the hybrid Metric-Palatini gravity through the GDE of the null vector fields. Furthermore, the obtained results indicate that the general performances of the observer area-distance and the null deviation vector fields in the hybrid Metric-Palatine gravity for a matter-dominated universe are almost similar to other corresponding modified gravity theories. We can summarize that the behavior of the deviation vector in the modified gravity theories at a low redshift regime is similar to the $\Lambda$CDM model, according to the principle of correspondence. This means that our results in these theories fluctuate around GR with small corrections like a cosmological constant.
To use the applications of this study, we can say that the equation of the area-distance (\ref{eq63}) and (\ref{eq64}) can be applied to compute the angular size versus redshift based on the  Sunyaev-Zel'dovich effect \cite{B,Y}, and to compact the radio sources as cosmic rulers \cite{01}. Additionally, by using the relation between the area and the luminosity distances \cite{02}, there are possibilities to extend studies of the GDE in hybrid Metric-Palatini gravity with the data obtained from the observations of SNIa \cite{91}.


\begin{thebibliography}{9}
\bibitem{1} C. Corda, Int J. Mod. Phys. D \textbf{18} (2009) 2275.
\bibitem{2} R. M. Wald,  {\it General Relativity}, The University of Chicago Press, Chicago, (1984).
\bibitem{3} A. Guarnizo, L. Castaeda and  J. M. Tejeiro, Gen. Rel. Grav. \textbf{43} (2011) 2713.
\bibitem{4} J. L. Synge, Gen. Rel. Grav. \textbf{41}  (1934) 1195.
\bibitem{4b} H-J. Schmidt. Class. Quant. Grav.\textbf{7} (1990) 1023.
\bibitem{4c} D. Wands. Class. Quant. Grav. \textbf{11} (1994) 269.
\bibitem{4d} G. J. Olmo, Int. J. Mod. Phys. D \textbf{20} (2011) 413.
\bibitem{4e} S. Capozziello, Int. J. Mod. Phys. D 11, 483 (2002);\\
 S. M. Carroll, V. Duvvuri, M. Trodden and M. S. Turner, Phys. Rev. D \textbf{70} (2004) 043528;\\
 S.~Nojiri and S.~D.~Odintsov, Phys. Rev. D \textbf{68} (2003) 123512, [arXiv:hep-th/0307288 [hep-th]];\\
 S.~Nojiri and S.~D.~Odintsov, Phys. Rept. \textbf{505} (2011) 59, [arXiv:1011.0544 [gr-qc]];\\
S.~Nojiri, S.~D.~Odintsov and V.~K.~Oikonomou, Phys. Rept. \textbf{692} (2017) 1, [arXiv:1705.11098 [gr-qc]].
\bibitem{4f} M. Ferraris, M. Francaviglia and I. Volovich, [arXiv:9303007[gr-qc]];\\
 D. N. Vollick, Phys. Rev. D \textbf{68} (2003) 063510;\\
E. E. Flanagan, Class. Quant. Grav. \textbf{21} (2003) 417;\\
X. H. Meng and P. Wang, Phys. Lett. B \textbf{584} (2004) 1 ;\\
B. Li and M. C. Chu, Phys. Rev. D \textbf{74} (2006) 104010 ;\\
T. P. Sotiriou and S. Liberati, Ann. Phys. \textbf{322} (2007) 935;\\
G.~Allemandi, A.~Borowiec, M.~Francaviglia and S.~D.~Odintsov, Phys. Rev. D \textbf{72} (2005) 063505, [arXiv:gr-qc/0504057 [gr-qc]].
\bibitem{4h} T. Harko, T. S. Koivisto, F. S. N. Lobo and G. J. Olmo, Phys. Rev. D \textbf{85} (2012) 084016.
\bibitem{4i} N. Tamanini and C. G. Boehmer, Phys. Rev. D \textbf{87} (2013) 084031, [arXiv:1302.2355v1[gr-qc]];\\
F. Bombacigno, F. Moretti and G. Montani, Phys. Rev. D \textbf{100} (2019) 124036;\\
J. L. Rosa, J. P. S. Lemos and F. S. N. Lobo, Phys. Rev. D \textbf{98} (2018) 064054;\\
J. L. Rosa, S. Carloni, J. P. S. Lemos and Francisco S. N. Lobo, Phys. Rev. D \textbf{95} (2017) 124035;\\ 
J. L. Rosa, S. Carloni and J. P. S. Lemos, Phys. Rev. D \textbf{101} (2020) 104056;\\ 
J. L. Rosa, J. P. S. Lemos and Francisco S. N. Lobo, Phys. Rev. D \textbf{101} (2020) 044055;\\ 
J. L. Rosa, Phys. Rev. D \textbf{104} (2021) 064002. 
\bibitem{4j} S. Capozziello, T. Harko, F. S. N. Lobo and G. J. Olmo, Int. J. Mod. Phys. D \textbf{22} (2013) 1342006;\\
J. L. Rosa, D. A. Ferreira, D. Bazeia and F. S. N. Lobo, Eur. Phys. J. C \textbf{81} (2021) 20;\\ 
J. L. Rosa, F. S. N. Lobo and D. Rubiera-Garcia, JCAP\textbf{07} (2021) 009;\\
J. L. Rosa and J. P. S. Lemos, Phys. Rev. D \textbf{104} (2021) 124076.
\bibitem{4k} S. Capozziello, T. Harko, T. S. Koivisto, F. S. N. Lobo and G. J. Olmo, JCAP 04 (2013) 011, [arXiv:1209.2895 [gr-qc]].
\bibitem{4l} C. G. Boehmer, F. S. N. Lobo and N. Tamanini, Phys. Rev. D\textbf{ 88} (2013) 104019.
\bibitem{4m} A. Guarnizo, L. Castaneda and J. M. Tejeiro, Gen. Rel. Grav. \textbf{43} (2011) 2713;\\
F. Darabi, M. Mousavi and K. Atazadeh, Phys. Rev. D \textbf{91} (2015) 084023.
\bibitem{harko} T. Harko and F. S. N. Lobo, Phys. Rev. D \textbf{86} (2012) 124034.
\bibitem{houn} T.~Harko, F.~S.~N.~Lobo, S.~Nojiri and S.~D.~Odintsov, Phys. Rev. D \textbf{84} (2011) 024020, [arXiv:1104.2669 [gr-qc]];\\
E. H. Baffou, M. J. S. Houndjo, M. E. Rodrigues, A. V. Kpadonou and J. Tossa, Chin. J. Phys. \textbf{55} (2017) 467.
\bibitem{shahab} J. -Z. Yang, S. Shahidi, T. Harko, S.-D. Liang, Eur. Phys. J. C \textbf{81} (2021) 111.
\bibitem{meraj}S. M. M. Rasouli and F. Shojai, Phys. Dark Univ. \textbf{32} (2021) 100781.
\bibitem{Q} J. -T. Beh, T. -H Loo, A. De, Chin. J. Phys. \textbf{77} (2022) 1551.
\bibitem{farhodi}R. Zaregonbadi, N. Saba and M. Farhoudi, Eur. Phys. J. C \textbf{82} (2022) 730.
\bibitem{jalal}S. M. M. Rasouli, A. F. Bahrehbakhsh, S. Jalalzadeh and M. Farhoudi, Eur. Phys. Lett. \textbf{87} (2009) 40006.
\bibitem{meraj2} S. M. M. Rasouli, M. Sakellariadou and P. V. Moniz, Phys. Dark Univ. \textbf{37} (2022) 101112.
\bibitem{6} S. Carloni, T. Koivisto and F. S. N. lobo, Phys. Rev. D \textbf{92} (2015) 064035.
\bibitem{7} G. F. R. Ellis and H. Van Elst, [arXiv:9709060v1[gr-qc]].
\bibitem{8} N. Tamanini and C. G. Bohmer, Phys. Rev. D \textbf{87} (2013) 084031.
\bibitem{111}P. Schneider, J. Ehlers and E. E. Falco, {\it Gravitational Lenses}, (Springer Verlag, Berlin, 1992).
\bibitem{9} A. L. Berkin and K. -I. Maeda, Phys. Rev. D \textbf{44}  (1991) 1691;\\
A. A. Starobinsky and J. i. Yokoyama, [arXiv:9502002[gr-qc]];\\
A. A. Starobinsky, S. Tsujikawa and J. i. Yokoyama, Nucl. Phys. B \textbf{610} (2001) 383.
\bibitem{10} J. Garcia-Bellido and D. Wands, Phys. Rev. D \textbf{52}  (1995) 6739;\\
J. Garcia-Bellido and D. Wands, Phys. Rev. D \textbf{53}  (1996) 5437;\\
F. Di Marco, F. Finelli and R. Brandenberger, Phys. Rev. D \textbf{67}  (2003) 063512;\\
F. Di Marco and F. Finelli, Phys. Rev. D \textbf{71}  (2005) 123502.
\bibitem{B}M. Bonamente, M. K. Joy, S. J. LaRoque, J. E. Carlstrom, E. D. Reese and K. S. Dawson, Astrophys. J. 647 (2006) 25.
\bibitem{Y}Y. Chen and B. Ratra, Astron. Astrophys. \textbf{543} (2012) A104.
\bibitem{01}J. A. S. Lima and J. S. Alcaniz, Astrophys. J. \textbf{566} (2002) 15;\\
 J. C. Jackson, Mon. Not. Roy. Astron. Soc. \textbf{390} (2008)  L1.
\bibitem{02}D. R. Matravers and A. M. Aziz, Mon. Not. Astron. Soc. South. Afr. \textbf{47} (1988) 124.
\bibitem{91}N. Suzuki {\it et al.}, Astrophys. J. \textbf{746} (2012) 85;\\
 H. Campbell {\it et al.}, Astrophys. J. \textbf{763} (2013) 88.
\end{thebibliography}
\end{document}